\documentclass[twocolumn, tighten, twocolappendix]{aastex631}

\usepackage{amsmath}
\usepackage{subfigure}
\usepackage{enumitem}

\begin{document}

\vspace*{-1.7cm}

\title{Modeling the imprints of large-scale magnetized structures on $\gamma$-rays from extragalactic transients}

\author[0000-0001-5796-225X]{Saikat Das}
\affiliation{Center for Gravitational Physics and Quantum Information, Yukawa Institute for Theoretical Physics, \\Kyoto University, Kyoto 606-8502, Japan}
\affiliation{Department of Physics, University of Florida, Gainesville, FL 32611, USA}
\email{saikatdas@ufl.edu}  

\author[0000-0002-0130-2460]{Soebur Razzaque} 
\affiliation{Centre for Astro-Particle Physics (CAPP) and Department of Physics, University of Johannesburg, \\PO Box 524, Auckland Park 2006, South Africa}
\affiliation{Department of Physics, The George Washington University, Washington, DC 20052, USA}
\affiliation{National Institute for Theoretical and Computational Sciences (NITheCS), Private Bag X1, Matieland, South Africa}
\email{srazzaque@uj.ac.za}

\author[0000-0002-7021-5838]{Nestor Mirabal}
\affiliation{Center for Space Sciences and Technology, University of Maryland, Baltimore County, Baltimore, MD 21250}
\affiliation{Mail Code 661, Astroparticle Physics Laboratory, NASA Goddard Space Flight Center, Greenbelt, MD 20771, USA}
\affiliation{Center for Research and Exploration in Space Science and Technology, NASA Goddard Space Flight Center, Greenbelt, MD 20771}
\email{nestorm@umbc.edu}

\author[0000-0002-5448-7577]{Nicola Omodei}
\affiliation{W. W. Hansen Experimental Physics Laboratory, Kavli Institute for Particle Astrophysics and Cosmology, \\Department of Physics and SLAC National Accelerator Laboratory, Stanford University, Stanford, CA 94305, USA}

\author[0000-0002-5358-5642]{Kohta Murase}
\affiliation{Department of Physics; Department of Astronomy \& Astrophysics; Center for Multimessenger Astrophysics, \\Institute for Gravitation and the Cosmos, The Pennsylvania State University, University Park, PA 16802, USA}
\affiliation{Center for Gravitational Physics and Quantum Information, Yukawa Institute for Theoretical Physics, \\Kyoto University, Kyoto 606-8502, Japan}

\author[0000-0002-2471-8696]{Israel Martinez-Castellanos}
\affiliation{Dept. of Physics, University of Maryland, College Park, MD 20742, USA}

\begin{abstract}

In the multimessenger era, the association of high-energy transients with their $\gamma$-ray counterparts can be obscured by magnetized large-scale structure surrounding the source. We analyze models for delayed off-axis $\gamma$-rays, arising from the injection of ultrahigh-energy cosmic rays ($E\gtrsim 10^{18}$ eV) and $\gamma$-rays ($\varepsilon_\gamma\gtrsim10^{18}$ eV), and very-high-energy $\gamma$-rays ($\varepsilon_\gamma\gtrsim 100$ GeV) from the transient. As a representative case, we consider the brightest gamma-ray burst, GRB~221009A, embedded in a magnetized structure. We examine charged-particle deflection in electromagnetic cascades, followed by propagation of secondaries through the intergalactic medium, and derive the time delays and angular offsets of the resulting $\gamma$-rays. To test the models, we analyze 1.25 years of post-event \textit{Fermi} Large Area Telescope (LAT) data and construct a 1 GeV -- 1 TeV test statistic (TS) map within $10^\circ$ of the burst localization. We find two sub-threshold peaks with TS$\geq9$. The more significant peak, J1911.8+2044, shows $\gamma$-ray emission in pre-burst LAT data, while J1913.2+1901 coincides with a 664.6 GeV $\gamma$-ray recorded $\sim191.9$ days after the trigger at $\approx 0^{\circ}.75$ from the GRB. The corresponding flux upper limit provides a fiducial value for constraining the parameter space in our models and estimating the source energy requirement for delayed off-axis signals. Although a magnetized structure can reproduce the angular offset of this sub-threshold feature, the resulting time delay disfavors an origin in this GRB.

\end{abstract}




\section{Introduction\label{sec:intro}} 

\begin{figure*}
    \centering
    \subfigure[]{
    \includegraphics[width=0.6\linewidth]{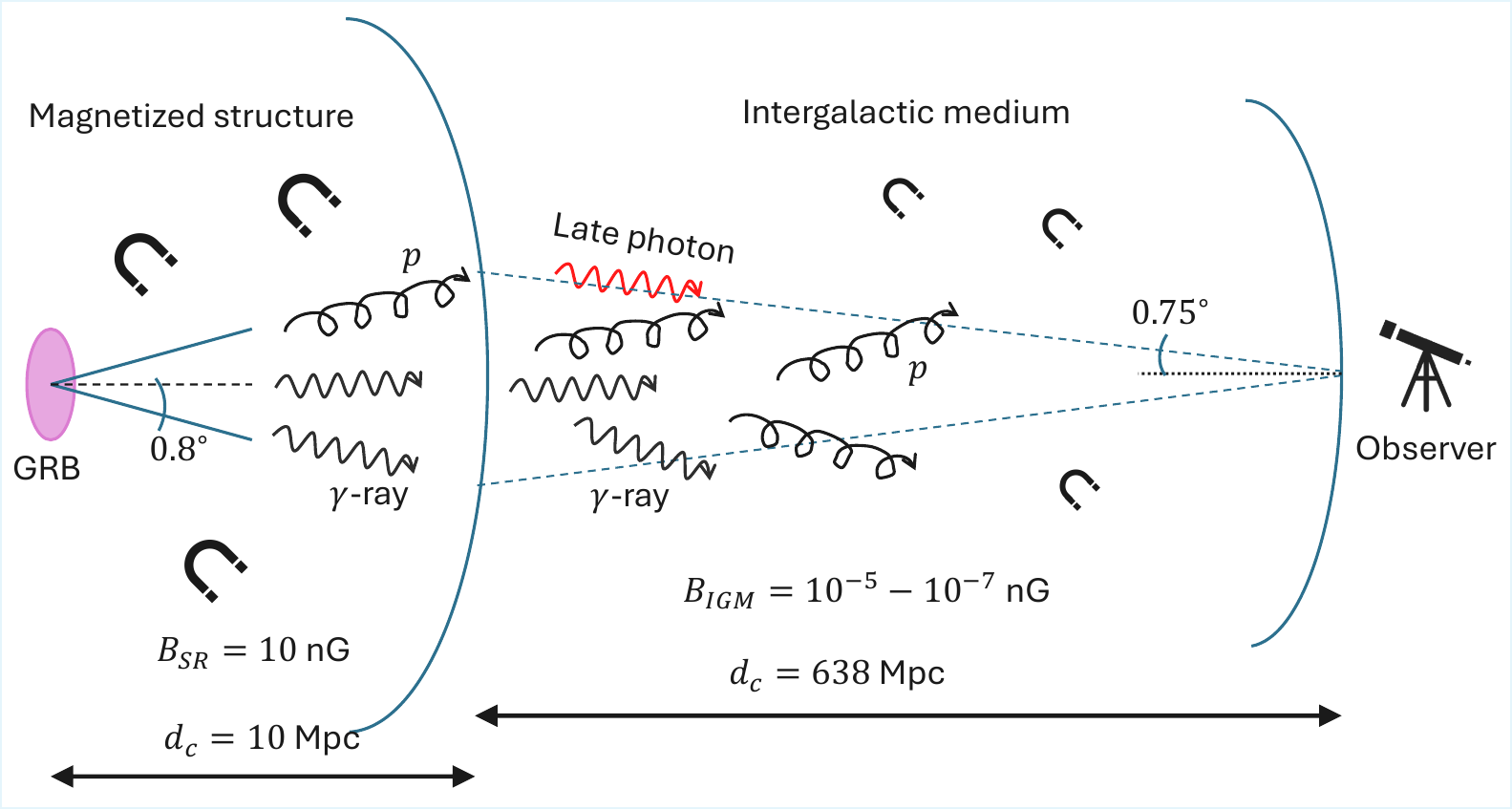}
    \label{fig:1a}
    }%
    \hfill
    \subfigure[]{
    \includegraphics[width=0.35\textwidth]{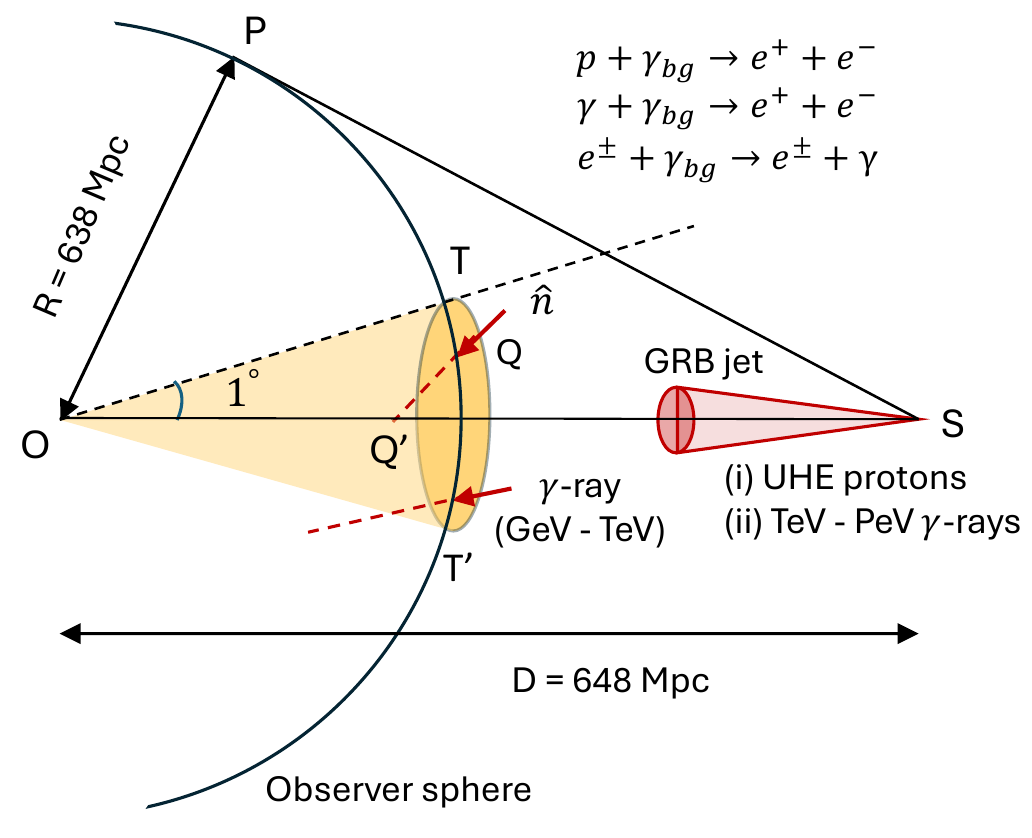}
    \label{fig:1b}
    }
    \caption{(a) Schematic illustration of the two distinct zones: the magnetized structure hosting the GRB at redshift $z=0.151$, and the surrounding void region. We adopt a comoving size of $d_c \sim 10$ Mpc for the structured region, while the weakly magnetized void IGM extends over $d_c \sim 638$ Mpc.
(b) The observer sphere ($R_{\rm obs}=638~{\rm Mpc}$) is centered at $O$, used to record cascade $\gamma$ rays produced after charged particles ($p$, $e^\pm$) are deflected within the magnetized structure. Events arriving within an aperture angle $\angle POS$ are collected. In this analysis, we apply a narrow angular cut defined by the yellow cone, with half-opening angle $\angle TOS = 1^\circ$ about the line of sight $OS$. The photon deflection angle is defined as the angle between the photon momentum vector $\hat{\mathbf{n}}$ (e.g., $\Vec{QQ'}$) and the initial jet emission axis, $\Vec{SO}$, taken to be $\hat{\mathbf{e}} = (-1,0,0)$. For a statistically significant event sample, the distribution around the line of sight may be regarded as azimuthally symmetric.
    }
    \label{fig:schematic}
\end{figure*}

High-energy $\gamma$-rays from transients can have an astrophysical origin, produced within the sources. Alternatively, they may have a cosmogenic origin due to the injection of ultrahigh-energy cosmic rays (UHECRs; $E\gtrsim10^{18}$ eV) and their subsequent interactions. 
For extragalactic, luminous $\gamma$-ray emitters, injected TeV-PeV $\gamma$-rays, UHE $\gamma$-rays, and UHECRs can interact with the cosmic background photons, namely the cosmic microwave background (CMB) and the extragalactic background light (EBL), which consists of infrared/optical/ultraviolet photons. The resulting high-energy secondary $e^\pm$ and $\gamma$-rays initiate an electromagnetic cascade down to GeV-TeV energies. 
Magnetized large-scale structures surrounding a source can leave imprints on the high-energy $\gamma$-ray signal by deflecting charged particles. In this work, we quantify the deflection and time delay in such scenarios using the representative case of the brightest gamma-ray burst, GRB 221009A \citep{2023ApJ...946L..31B} at redshift $z=0.151$, modeled as embedded in a galaxy cluster region.

The rising afterglow emission and the highest energy $\gamma$-rays from a GRB were observed from GRB~221009A, due to the source being within the field of view of the Large High Altitude Air Shower Observatory \citep[LHAASO;][]{LHAASO:2023kyg, LHAASO:2023lkv}. LHAASO detected $\gamma$-rays up to 13 TeV, with more than 140 photons above 3 TeV. These photons arrived 230-900~s after the \textit{Fermi} Gamma-Ray Burst Monitor (GBM) trigger 
\citep[$T_0$;][]{2023ApJ...952L..42L}. 
\textit{Fermi}-GBM has observed two distinct emission characteristics. An isolated peak between $T_0$ and $T_0+20$~s was observed by GBM and other $\gamma$-ray detectors, corresponding to a highly variable MeV emission. This was followed by an extended emission over $T_0+220$~s to $T_0+550$~s coinciding in time with the very-high-energy (VHE; $\varepsilon_\gamma\gtrsim 100$ GeV) $\gamma$-ray flux, with a relatively smooth temporal profile. Simultaneously, the {\it Fermi} Large Area Telescope ({\it Fermi}-LAT) detected the highest-energy $\gamma$-ray from a GRB of energy 99.3 GeV, 240~s after the trigger \citep{Bissaldi_2022, Pillera_2022, Axelsson_2025}. In addition, {\it Fermi}-LAT detected a 400 GeV photon 33~ks after the trigger, which is incompatible with a purely leptonic or an associated TeV electromagnetic cascade emission \citep{Axelsson_2025}, indicating a possible hadronic origin. 

The air-shower array Carpet-3 detected a photon of energy $300^{+43}_{-38}$ TeV from the direction of this GRB at $T_0+4536$~s \citep{Carpet-3Group:2025fcs}. If its association with~GRB 221009A is real, such a photon would require physics beyond standard EBL/CMB attenuation, such as photon--axion-like particle (ALP) conversion \citep{Galanti:2022chk, Galanti:2022xok, Zhang:2022zbm, Baktash:2022gnf, Troitsky:2022xso, Nakagawa:2022wwm, Carenza:2022kjt} or Lorentz Invariance violation \citep[LIV;][]{LHAASO:2024lub, Hua:2025cda, Li:2023rhj}. First-order subluminal LIV could reduce the opacity \citep{Finke:2022swf}, although the observed delay introduces additional constraints, including on higher-order LIV terms \citep{Ofengeim:2025jsw}, and intrinsic source lags remain possible \citep{Vardanyan:2022ujc}.

The unusual characteristics of the GRB~221009A event have provided critical insight into GRB physics and the origin of high-energy cosmic rays, $\gamma$ rays, and neutrinos \citep{2022ApJ...941L..10M}.  Recent studies have proposed that $\gtrsim 10$ TeV $\gamma$ rays detected by LHAASO from GRB 221009A may arise from the UHECR-induced secondary emission in the intergalactic medium \citep{2022arXiv221012855A, Das:2022gon, 2023MNRAS.519L..85M}, providing a possible cosmogenic origin. Alternatively, our astrophysical scenarios consider the unattenuated
VHE synchrotron self-Compton (SSC) emission and UHE $\gamma$-ray emission from the GRB jet, cascade down to GeV--TeV energies during
propagation through magnetized and the void intergalactic
medium (IGM).

The analysis presented here is crucial for interpreting time delays in observed $\gamma$-ray signals due to cosmic-ray propagation in magnetized cosmological volumes \citep[e.g.,][]{Murase:2011yw}. Large-scale structures
surrounding the GRB~221009A host galaxy may correspond to galaxy clusters,
filaments, or magnetized voids. Recent studies using LHAASO data indicate that
the $\gamma\gamma$ opacity at TeV energies can be significantly reduced for
intervening cosmic voids along the line of sight to GRB~221009A
\citep{2023MNRAS.519L..85M, Abdalla2024MNRAS.532..198A}. In our study, the structured region surrounding the GRB and the intergalactic medium differ only in magnetic-field strength and turbulence coherence length. We assume a highly
magnetized $\sim10~\mathrm{Mpc}$ region around the GRB, followed by propagation
through the void IGM, as illustrated schematically in Fig.~\ref{fig:1a}. 
For the UHECR and VHE $\gamma$-ray injection scenarios, we perform fully 3-D
simulations of particle propagation and secondary electromagnetic cascades, and
show that the jet-opening angle is crucial for delayed off-axis emission. The UHE $\gamma$-ray injection scenario, which produces pair-echo synchrotron
emission, is treated semi-analytically, since a self-consistent numerical calculation
requires dedicated time-dependent cascade simulations beyond the scope of the
present work. 

To estimate the source energy requirements in all cases and to compare them with those in the GRB blastwave, we use fiducial flux values from a search for a delayed $\gamma$-ray emission around GRB 221009A.
We analyze 1 GeV--1 TeV data from {\it Fermi}-LAT between 5.73 to 456.9 days after the burst, once the initial GRB $\gamma$-ray emission has subsided \citep{Axelsson_2025}. Using a
$10^\circ\times10^\circ$ region centered on the burst localization, we identify
a sub-threshold feature associated with a 664.6 GeV photon at an angular offset
of $0^\circ.75$ and $\sim191.9$ days post-trigger, with a statistical
significance of $3\sigma$. We show that magnetized-structure and intergalactic
magnetic field (IGMF) can account for angular offsets of this order for UHECR and VHE $\gamma$-ray injection scenario.
In addition, we show that synchrotron
emission from pairs produced by UHE $\gamma$-rays at $\sim10^{19}$ eV in the structured region can also yield delayed emission 
although the emission is highly beamed.
However, the geometric time delay implied by the observed offset already exceeds the observed delay for a source at the distance of GRB~221009A, disfavouring a
physical association with this burst.
The resulting limits provide a benchmark for the detectability of
delayed off-axis cascade $\gamma$-ray signals from past high-energy transients.

In Sec.~\ref{sec:models}, the numerical analysis of our models is presented. In Sec.~\ref{sec:fermi}, we present the Fermi-LAT analysis of the GeV-TeV emission surrounding the GRB and Sec.~\ref{subsec:broadband} tests the consistency with the models studied. We discuss the results and draw our conclusions in Sec.~\ref{sec:discussion}. Throughout the paper, we assume a $\Lambda$CDM cosmology with dimensionless Hubble parameter $h=0.673$, 
$\Omega_{m} = 0.315$, and $\Omega_{\Lambda} = 0.685$.


\section{Models of Delayed Off-axis Emission\label{sec:models}}

\subsection{Analytical Estimates\label{subsec:uhecr}}
UHE protons escaping from the GRB jet are injected in the magnetized volume surrounding the source. We assume a conservative value of the RMS magnetic field strength $B_{\rm SR} = 10$~nG and coherence length, $\lambda_c=10$ kpc for a galaxy cluster region \citep{Medina-Tanco:2001qmg, Carilli2002ARA&A..40..319C,  Marinacci:2017wew} of length scale $d_c\approx 10$~Mpc. 
The deflection angle for a UHECR with charge $Z$ and gyroradius $R_L$ is $\theta_{\rm dfl}= {\sqrt{2d_c\lambda_c}}/{3R_L}$ and can be written as \citep{Kotera:2008ae, Dermer2009NJPh}
\begin{equation}
\begin{split}
    \theta_{\rm dfl} \simeq 0^\circ.75 \ Z 
    & \left( \frac{d_c}{10~\rm Mpc} \right)^{1/2}
     \left( \frac{\lambda_c}{10~\rm kpc} \right)^{1/2} \nonumber \\ 
    &\times\left( \frac{B_{\rm SR}}{10~\rm nG} \right)  \left( \frac{\varepsilon_{{\rm UHE}p}}{100~\rm EeV} \right)^{-1}\,
    \label{eqn:theta_dfl}
   \end{split}
\end{equation}
The time delay induced by the deflection can be expressed as \citep{Murase:2011cy, Takami:2011nn},
\begin{equation}
\begin{split}
\Delta t_{\rm SR}& \simeq \frac{d_c \theta_{\rm dfl}^2}{2c} \simeq 2.8  \times10^3~\text{yrs} \times\bigg(\dfrac{d_c}{10\text{ Mpc}}\bigg)^{2} \nonumber \\
&\times\bigg(\dfrac{\lambda_c}{10\text{ kpc}}\bigg)\bigg(\dfrac{B_{\rm SR}}{10\text{ nG}}\bigg)^2 \bigg(\dfrac{\varepsilon_{{\rm UHE}p}}{100~\text{EeV}}\bigg)^{-2}
\label{eqn:delay_igm} \,.
\end{split}
\end{equation}
Deflection of UHECRs in the IGM in the void region is negligible for current constraints on magnetic field, $B_{\rm IGM} \sim 10^{-14} - 10^{-16}$~G \citep{2010Sci...328...73N, 2023ApJ...950L..16A}, although secondary $e^\pm$ from electromagnetic cascade can be deflected away from the line of sight depending on the strength. Recent studies of anisotropic GeV pair halos around stacked TeV blazars also highlight the role of magnetic deflection of secondary pairs, with an inferred EGMF strength of $\sim3\times10^{-16}\,\mathrm{G}$ \citep{Zhang:2026lte}. 
The geometric time delay of off-axis photons relative to those coming directly from the GRB is (see Fig.~\ref{fig:schematic}),
\begin{equation}
\begin{split}
     \Delta t_{g} &= (\sqrt{D^{\prime 2} + D^2 - 2DD^\prime \cos\theta_{\rm dfl}} + D^\prime - D)/c \label{eqn:time_geo} \\
     &\simeq 2.6\times10^7\,\text{yr} \nonumber
     \end{split}
\end{equation}
assuming $D^\prime\approx D$, where $D=604$ Mpc is the light travel distance 
to the GRB and $D^\prime$ is the distance to the secondary source at a transverse distance due to an angular offset of $\theta_{\rm dfl} =0^\circ.75$, respectively, from the observer. 

The analytic expression for the additional time delay due to the electromagnetic cascade emission in the void IGM depends on the inverse Compton (IC) cooling time scale of $e^\pm$ pairs on the CMB photons, $t^\prime_{\rm IC} = \lambda^\prime_{\rm IC}/c = 3m_ec/(4\gamma_e \sigma_{\rm T} u^\prime_{\rm CMB}) \simeq 1.3\times 10^{5}(\gamma_e/10^7)^{-1}$~yr, where $\lambda^\prime_{\rm IC}$ is the local IC cooling length in terms of the local CMB energy density $u^\prime_{\rm CMB}$ at a redshift $z=0.151$, and $\gamma_e$ is the electron Lorentz factor \citep[see, e.g.,][]{Dai2002ApJ...580.1013D, Razzaque:2004ApJ...613.1072R, Murase:2007ar}.
%
\begin{figure*}
    \centering
    \subfigure[$\gamma$-rays]{
    \includegraphics[width=0.48\textwidth]{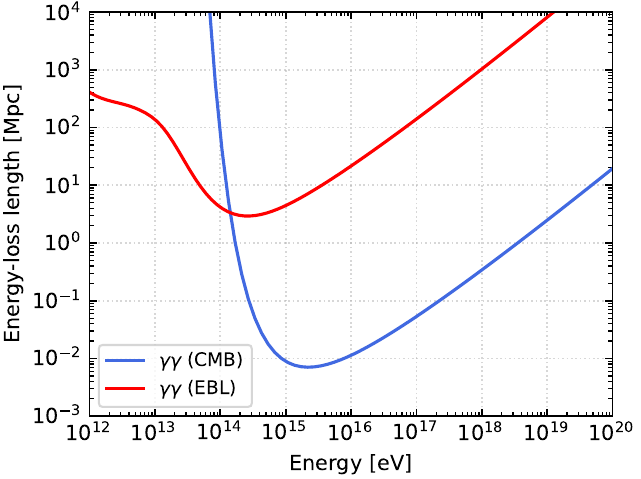}
    }
    \hfill
    \subfigure[Electrons]{
    \includegraphics[width=0.48\textwidth]{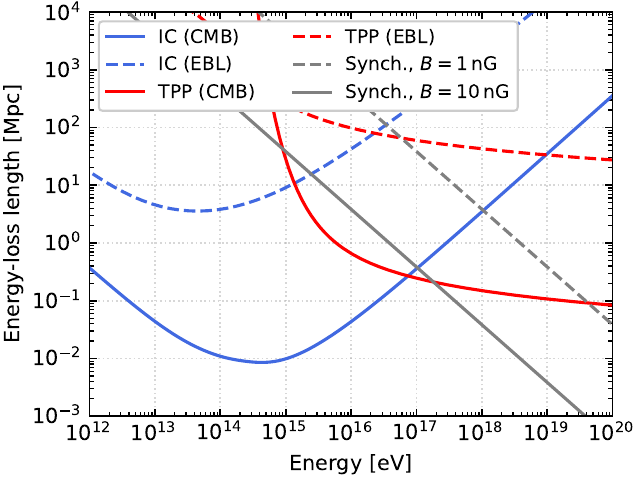}
    }
    \caption{Energy-loss lengths of (a) $\gamma$-rays and (b) electrons, interacting with CMB and EBL \citep{Gilmore_2012} photons. Note that the interaction length for IC scattering differs from the energy-loss length $\lambda_{\rm IC}= c\gamma_e/|\dot{\gamma}_e|$ shown here. For triplet-pair production, we assume a conservative inelasticity of $\eta=0.5$. The grey dashed and solid lines correspond to synchrotron radiation in magnetic field strengths of 1 and 10 nG. $\gamma$-ray attenuation with the cosmic radio background occurs at $\gtrsim 10^{20}$ eV. IC losses on the radio background \citep{Protheroe:1996si} are negligible compared with IC losses on the CMB and synchrotron losses for the parameter range considered here.
    }
    \label{fig:loss_length}
\end{figure*}

The deflection of the pairs in the IGM is calculated from the ratio between the IC path length and the Larmor radius $R_{\rm L} = \gamma_e m_ec^2/eB_{\rm IGM}$. 
For sufficiently small deflections in weak IGMF, the magnetic deflection angle is given by 
\begin{equation}
\begin{split}
    \theta_B \approx \frac{ct^\prime_{\rm IC}}{R_{\rm L}} \simeq 4.2^\circ \left(\frac{\gamma_e}{10^7}\right)^{-2} \left(\frac{B_{\rm IGM}}{10^{-14}~{\rm G}}\right).
    \label{eqn:theta_cascade}
    \end{split}
\end{equation} 
For a pencil beam, the time delay due to the bending of the electron trajectory in the magnetic field is 
$\Delta t^\prime_{B} \approx t^\prime_{\rm IC} \theta_{B}^2/2 \simeq 360~\left({\gamma_e}/{10^7}\right)^{-5} \left({B_{\rm IGM}}/{10^{-14}~{\rm G}}\right)^{2}$~{yrs}.

However, in a more accurate treatment, the angular spreading time is given by $\Delta t^\prime_B=(\lambda^\prime_{\rm IC}+\lambda^\prime_{\gamma\gamma})\theta_B^2/2c$. The delay due to $\gamma\gamma$ pair production can be parametrized as $\approx \lambda^\prime_{\gamma\gamma} \theta_{B}^2/(2c) \simeq 3.5\times 10^5 (\gamma_e/10^7)^{-4} (B_{\rm IGM}/10^{-14}~{\rm G})^{2}$~yrs \citep{Ichiki08, Murase08, Murase09}.
For $\gamma$-rays with energy $\varepsilon_\gamma\gtrsim1$ TeV and approximating $\sigma_{\gamma\gamma}\simeq0.26\sigma_T$, the local $\gamma\gamma$ mean free path in terms of the local cosmic infrared background (CIB) photon density $n^\prime_{\rm CIB}$ is given as $\lambda^\prime_{\gamma\gamma}\approx 190\, \text{Mpc} \times (n^\prime_{\rm CIB}/0.01\text{ cm}^{-3})^{-1}$. The mean free path decreases with increasing $\gamma$-ray energy, reaching $\sim$few Mpc at $\sim100$~TeV for CIB interactions, while CMB absorption dominates in the PeV range \citep{Heiter:2017cev}.
We show the relevant energy-loss and interaction lengths for secondary electrons and $\gamma$-rays with CMB and EBL photons in Fig.~\ref{fig:loss_length}. Synchrotron losses of secondary electrons are negligible in the UHECR and VHE $\gamma$-ray injection cases, even in the structured-region magnetic field, but dominate for the UHE $\gamma$-ray injection case.


\subsection{Simulation Setup\label{subsec:setup}}

\begin{figure*}
\centering
    \subfigure[]{
    \includegraphics[width=0.31\textwidth]{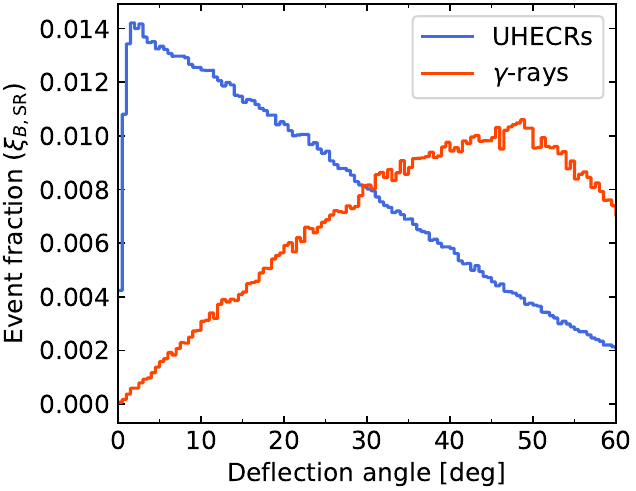}
    \label{fig:uhecr_defl}
    }%
    \subfigure[]{
    \includegraphics[width=0.33\textwidth]{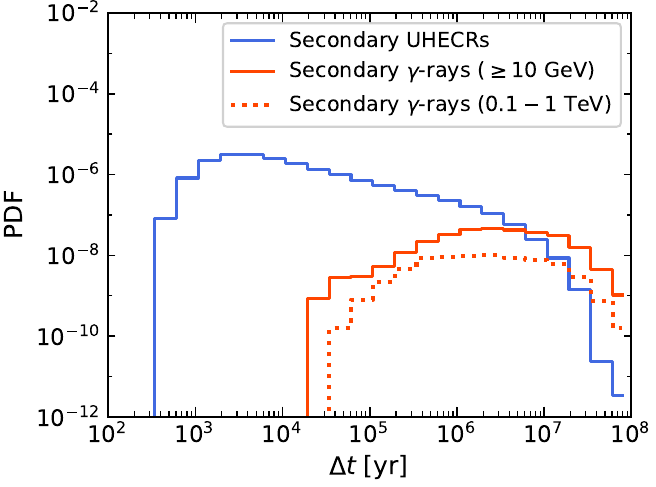}
    \label{fig:uhecr_delay}
    }%
    \subfigure[]{
    \includegraphics[width=0.315\textwidth]{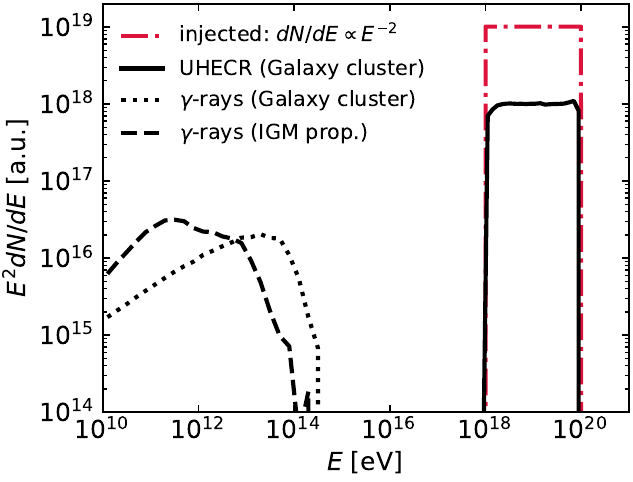}
    \label{fig:flux_uhecr}
    }
    \caption{{UHECR ($E_p \sim 10^{18}-10^{20}$ eV) injection model} --- (a) Distribution of the angle between the initial emission axis and the momentum vector of secondary UHECR and cascade photons at $\varepsilon_\gamma>10$ GeV, after propagation in the magnetized region. (b) PDF of the time delay distribution of secondary UHECRs and $\gamma$-rays emerging from the magnetized region with respect to the rectilinear travel time. (c) The solid line shows the secondary UHE-proton spectrum after propagation through the galaxy-cluster region, and the dotted line shows the associated cascade $\gamma$-ray flux. Further propagation through the IGM produces the cosmogenic $\gamma$-ray flux at Earth, shown by the dashed line. The fluxes are normalized to 10 injected protons.
    }
\end{figure*}
%


We use the Monte Carlo code {CRPropa 3.2}  \citep{AlvesBatista:2016vpy, AlvesBatista:2022vem} to perform 3-D simulations of UHECR and VHE $\gamma$-ray propagation. Since the propagation delays considered here are much longer than the intrinsic
GRB emission duration, we treat the UHECR and VHE $\gamma$-ray source
injections as instantaneous.
For the galaxy cluster region with $d_c=10$ Mpc, $B_{\rm SR}= 10$ nG, and $\lambda_c=10$ kpc, we set up a turbulent magnetic field with a Kolmogorov power spectrum over length scales of $20-25$ kpc. The field is stored on a $512^3$ grid with 10 kpc grid spacing. 
Primary particles are injected within a cone of half-opening angle $0.8^\circ$, corresponding to the GRB jet opening angle $\theta_j$ as predicted by the LHAASO collaboration \citep{LHAASO:2023kyg}. An observer sphere of radius 638 Mpc is used to collect the secondary $\gamma$-ray and UHECR events after propagation in the galaxy cluster, and events within an aperture angle of $1^\circ$ are kept for further propagation through IGM, as shown in Fig.~\ref{fig:1b}.
We similarly model the IGMF by a random turbulent magnetic field with a Kolmogorov power spectrum over $0.1-5$~Mpc length scales, $\lambda_c=1$ Mpc, and is stored on a 256$^3$ grid with 30 kpc grid spacing. 
We show in the following sections that our choice of (i) $B_{\rm IGM} \simeq 10^{-7}$~nG for the UHE proton injection case and (ii) $B_{\rm IGM}\simeq10^{-5}$ nG for the VHE $\gamma$-ray injection case, facilitates the observation of a significant fraction of the secondary cosmogenic $\gamma$ rays from the source. Similar to the galaxy cluster region, events are selected within a $1^\circ$ aperture around the line of sight at the observer position.

We include all energy-loss processes for UHE protons, viz., photopion and Bethe-Heitler interactions with CMB and EBL \citep{Gilmore_2012}, and nuclear $\beta$-decay of secondary neutrons. The electromagnetic cascade includes pair production, double- and triplet-pair production, IC scattering of background photons, and synchrotron radiation in cosmic magnetic fields. 
The adiabatic energy loss due to the expansion of the Universe is also included for both UHECRs and all secondaries. In our simulations, the EBL intensity is implemented only as a function of photon energy and redshift, without spatial correlations with large-scale structure.
For computational efficiency, we calculate the cascade photon flux down to 10 GeV. At lower energies, uncertainties in the flux arise from discrepancies in energy conservation and various numerical methods employed \citep[see, e.g.,][]{Murase:2015xka, Kalashev:2022cja}.
For the UHECR- and VHE $\gamma$-ray-induced cascades in the magnetized
structure, synchrotron losses of secondary pairs are subdominant. For the UHE $\gamma$-ray injection scenario, however, pairs produced
in the structured region can radiate efficiently through synchrotron emission
for typical GRB~221009A parameters.

\subsection{UHE proton-induced Cascade Emission\label{subsec:cosmogenic}}

\subsubsection{Propagation in magnetized structured region\label{subsec:proton_propagation}}

We inject UHECR protons from the GRB jet with $dN/dE_p\propto E_p^{-2}$ spectrum in the 1-100 EeV energy range shown by the dashed-dotted curve in Fig.~\ref{fig:flux_uhecr}.
The cascade $\gamma$-ray spectrum and the secondary UHECR spectrum of events kept within the $1^\circ$ viewing aperture (see Fig.~\ref{fig:1b}) emerging from the galaxy cluster region are shown by the dotted curve and solid curve, respectively.
The spectra are integrated over propagation time corresponding to this distance, and the injection time period does not affect these energy fractions. The fluxes are presented in arbitrary units, normalized to 10 UHE protons injected from the GRB jet. The $\gamma$-ray flux falls off sharply beyond $\mathcal{O}(100)$ TeV.

Fig.~\ref{fig:uhecr_defl} shows the angular distribution of the momentum vectors of the kept events relative to the emission axis, binned in intervals of $0.5^\circ$. We denote this fraction by $\xi_{B,\rm SR}$.  The deflection for UHECRs peaks at around $1^\circ.5$, comparable to that from Eqn.~\ref{eqn:theta_dfl}, while that of secondary $\gamma$-rays from EM cascade peak at a much larger angle. The total fraction of events within $1^\circ$ is $\xi_{B, \rm SR}(\leq 1^\circ)=0.015$. For a power-law injection, the IC cascade component from lower-energy
pairs is significantly deflected away from the line of sight in the highly
magnetized galaxy-cluster region, as expected from Eqn.~\ref{eqn:theta_cascade},
and is therefore not propagated further through the IGM. 
Fig.~\ref{fig:uhecr_delay} shows the PDF ($dN/d\Delta t$) of the time delay $\Delta t$ for UHECRs and cascade $\gamma$-rays in the galaxy-cluster region. The time-delay distribution of photons in the energy range $0.1-1$ TeV is shown by a dotted line.
$\Delta t$ is measured as the time required along the line of sight from the source to the observer, relative to the trajectory distance. The corresponding peaks in the obtained delay distribution conform to the angular deflections, $\Delta t\simeq d\theta_d^2/2c$.



\begin{figure*}
    \centering
    \subfigure[]{
    \includegraphics[width=0.32\textwidth]{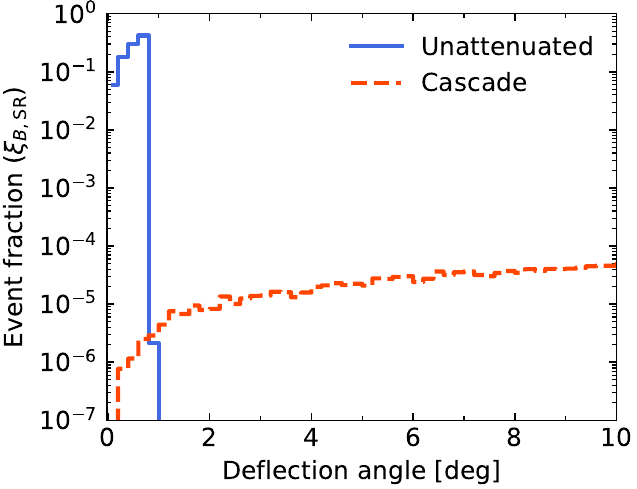}
    \label{fig:gamma_defl}
    }%
    \subfigure[]{
    \includegraphics[width=0.33\textwidth]{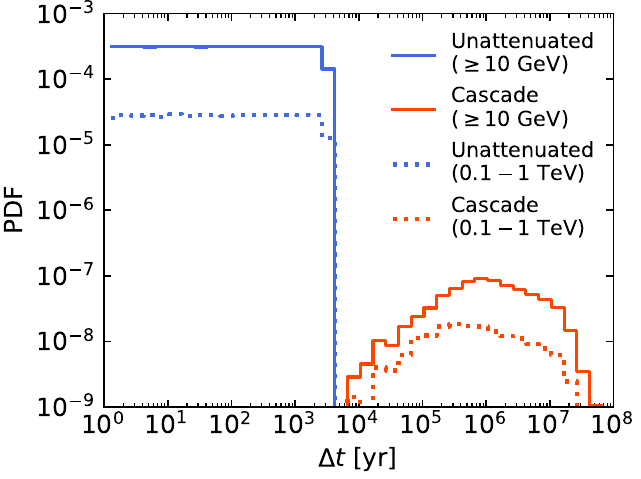}
    \label{fig:gamma_delay}
    }%
    \subfigure[]{
    \includegraphics[width=0.325\textwidth]{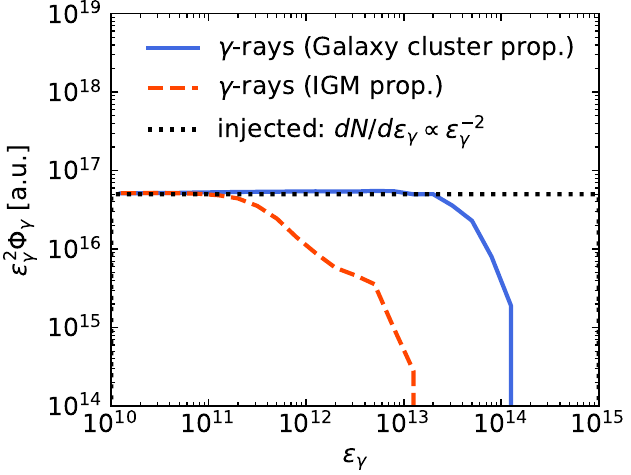}
    \label{fig:flux_gamma}
    }
    \caption{{VHE $\gamma$-ray ($\varepsilon_\gamma\sim 10^{-2} - 10^3$ TeV) injection model} --- (a) Distribution of the angle between the initial emission axis and the momentum vector angle of surviving unattenuated photons and secondary cascade photons at $\varepsilon_\gamma>10$ GeV, after propagation in the magnetized region. (b) PDF of the time-delay distribution of secondary photons emerging from the magnetized region with respect to the rectilinear travel time. (c) The blue solid line shows the secondary $\gamma$-ray spectrum after propagation through the magnetized structure, while the red dashed line shows the $\gamma$-ray spectrum at Earth 
    The fluxes are normalized to one injected $\gamma$-ray photon. 
    In both panels, spectra are shown within a $1^\circ$ viewing aperture at the structure boundary and at Earth.
    }
\end{figure*}

\subsubsection{\label{subsubsec:IGM}Propagation in Void IGM region}

We further simulate the IGM propagation of the secondary UHECR spectra produced in the magnetized structure region within $1^\circ$ aperture. However, we treat the secondary injection as being directed toward the observer, as if it originated from a point source. This choice is justified since for sufficiently low magnetic field strength, further deflection of the secondary $\gamma$-ray flux is negligible. We find that for $B_{\rm IGM}=10^{-7}$ nG, $\sim 50\%$ of the events on the surface of an observer sphere of radius $10$~Mpc (centered at Earth) lie within an aperture angle of $1^\circ$ around the line of sight, and the corresponding spectrum of these events is shown by the dashed line in Fig.~\ref{fig:flux_uhecr}. The energy fraction of the injected spectrum (dashed dotted curve) that goes into the observed spectrum (black dashed curve) is $f_{\gamma,p}\simeq0.0033$. We note that most secondary production occurs well beyond the observer sphere, as
shown by \citet{Das:2019gtu}. The momentum vectors of the $\gamma$-ray events observed within this aperture
remain closely aligned with the line of sight. Therefore, if the injection
follows the angular distribution at the edge of the galaxy cluster, the
$\gamma$-ray distribution within a $1^\circ$ aperture on the observer sphere is
approximately preserved. The momentum vectors of UHECR events at the edge of the galaxy cluster may point either toward or away from the line of sight at different locations on the surface of the aperture cone, as illustrated by the representative red dashed lines in Fig.~\ref{fig:1b}. Some of these events may give rise to a delayed off-axis $\gamma$-ray signal at Earth. Hence, a lower value of $B_{\rm IGM}\sim 10^{-7}$ nG is assumed in this case to minimize additional deflection.

\subsection{VHE $\gamma$-ray-induced Cascade Emission\label{subsec:gamma}}

%
%
Next, we explore the model in which $\gamma$-rays from the GRB jet are injected into the magnetized structure. 
We consider GeV-PeV $\gamma$ rays as the injected primaries from VHE emission produced in the 
afterglow.
As we will show later in our SSC modeling of the LHAASO data in Sec.~\ref{subsec:broadband}, a $dn/d\varepsilon_\gamma \propto\varepsilon_\gamma^{-2}$ spectrum can be used in the $10^{-2}$--$10^3$ TeV range for our simulation, as an approximation to the EBL-unattenuated SSC $\gamma$-ray flux. 
The $\gamma$-ray flux at the edge of the galaxy cluster is decomposed into an unattenuated component and a cascade component from $\gamma\gamma$ pair production followed by IC scattering, after selecting events within the $1^\circ$ aperture cone shown in Fig.~\ref{fig:1b}. At $100$~TeV, the $\gamma\gamma$ mean free path on the EBL is $\approx5$~Mpc, allowing pair production within the $10$~Mpc
cluster region. Fig.~\ref{fig:gamma_defl} shows the angular distributions of the momentum vectors of the selected events relative to the emission axis, using $0^\circ.2$ bins. The unattenuated $\gamma$-rays from the source preserve the uniform angular distribution per unit solid angle of the $0.8^\circ$ conical injection. Most IC $\gamma$ rays at $\gtrsim10$ GeV are scattered outside the $1^\circ$ aperture, with only a negligible fraction retained compared with the unattenuated component.
Fig.~\ref{fig:gamma_delay} shows the time-delay PDF, $dN/d\Delta t$, for the unattenuated and IC components.
The unattenuated component is nearly flat up to a few times $\sim 10^3$ yr, as expected for uniform injection over the jet solid angle, whereas the IC component peaks
at much later times, $\sim10^6$ yr. Because of their negligible retained fraction
and large deflection angles, IC events are neglected in the subsequent IGM
propagation. The delay distribution for photons in the energy range 0.1-1 TeV is shown by dotted lines for both components. The spectrum of the selected events at the edge of the cluster is
shown by the solid line in Fig.~\ref{fig:flux_gamma}.



%
The primary $\gamma$-rays escaping $\gamma\gamma$ pair production inside the magnetized structure region
are injected into the IGM. They produce $e^\pm$ pairs with CMB and EBL photons
which in turn can IC scatter the background photons.
This is generally known as ``inverse-Compton pair echo'' emission \citep{Murase08} as this cascade emission sustains for a longer duration than injection at source due to deflection of the $e^+e^-$ pairs in the IGMF in the void region, even after the original emission from the GRB ceases \citep{Plaga1995Natur.374..430P, Razzaque:2004ApJ...613.1072R, Murase:2007ar, Murase09, Murase:2009ah, Ichiki08, Takahashi2008ApJ...687L...5T}. 
Detection of such delayed emission can probe the strength of the IGMF in the void. Conversely, its non-detection from GRB~221009A implies a lower bound of $B_{\rm IGM}$ ranging from $\gtrsim 10^{-16}- 10^{-19}$~G. \citep{Huang_2023, Dzhatdoev_2024, Vovk2024A&A...683A..25V, Kalashev:2025slq, Burmeister:2025lgo}. 

However, at the edge of the galaxy cluster, the momentum vectors of the
unattenuated $\gamma$ rays point away from the line of sight due to the angular spread of the jet opening. Additional deflection of secondary $e^\pm$ pairs in the IGM is required to redirect the IC-scattered $\gamma$ rays toward Earth. We consider an RMS value of IGMF strength
$B_{\rm IGM}=10^{-5}$~nG, for which Eqn.~\ref{eqn:theta_cascade} gives the
cascade deflection. As before, we approximate the emission from the edge of the
galaxy cluster as a point source directed toward Earth. After selecting events within
a $1^\circ$ aperture cone on the observer sphere, we obtain the secondary spectrum
shown by the dashed line in Fig.~\ref{fig:flux_gamma}. The fraction of the injected VHE
$\gamma$-ray energy that goes to the observed spectrum is
$f_{\gamma,\gamma}=0.33$. This result constrains the IGMF required for VHE
$\gamma$-ray injection to produce a delayed off-axis signal.

%

\subsection{UHE $\gamma$-ray induced pair-echo emission}

\begin{figure}
    \includegraphics[width=0.47\textwidth]{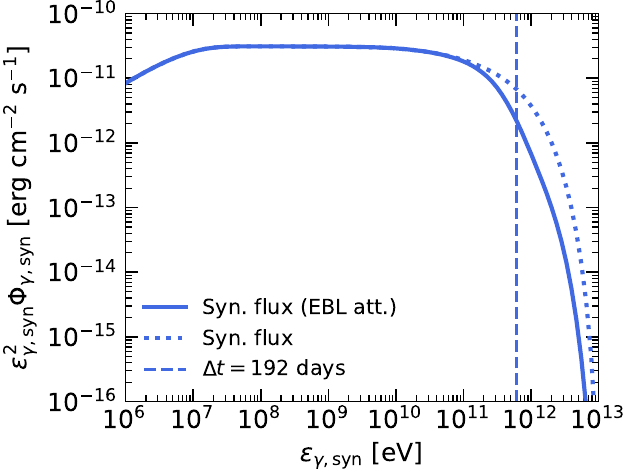}
    \caption{UHE $\gamma$-ray ($\varepsilon_\gamma\gtrsim 10^{18}$ eV) injection model--- Pair-echo synchrotron emission flux assuming $L_{\gamma,\rm iso}=10^{54}$ erg s$^{-1}$, $E_{p,\rm iso}\simeq6\times10^{51}$ erg, $f_{\rm esc}\rightarrow 1$. The dashed vertical line corresponds to the characteristic energy of synchrotron radiation at a time $\Delta t\simeq192$ days after the initial trigger. 
    The dashed vertical line marks the synchrotron photon energy for which the characteristic pair-echo time spread is $\Delta t_{\rm syn}\simeq192$ days.
    }
    \label{fig:pair_echo}
\end{figure}

\begin{figure*}
     \centering
    \includegraphics[width=0.48\textwidth]{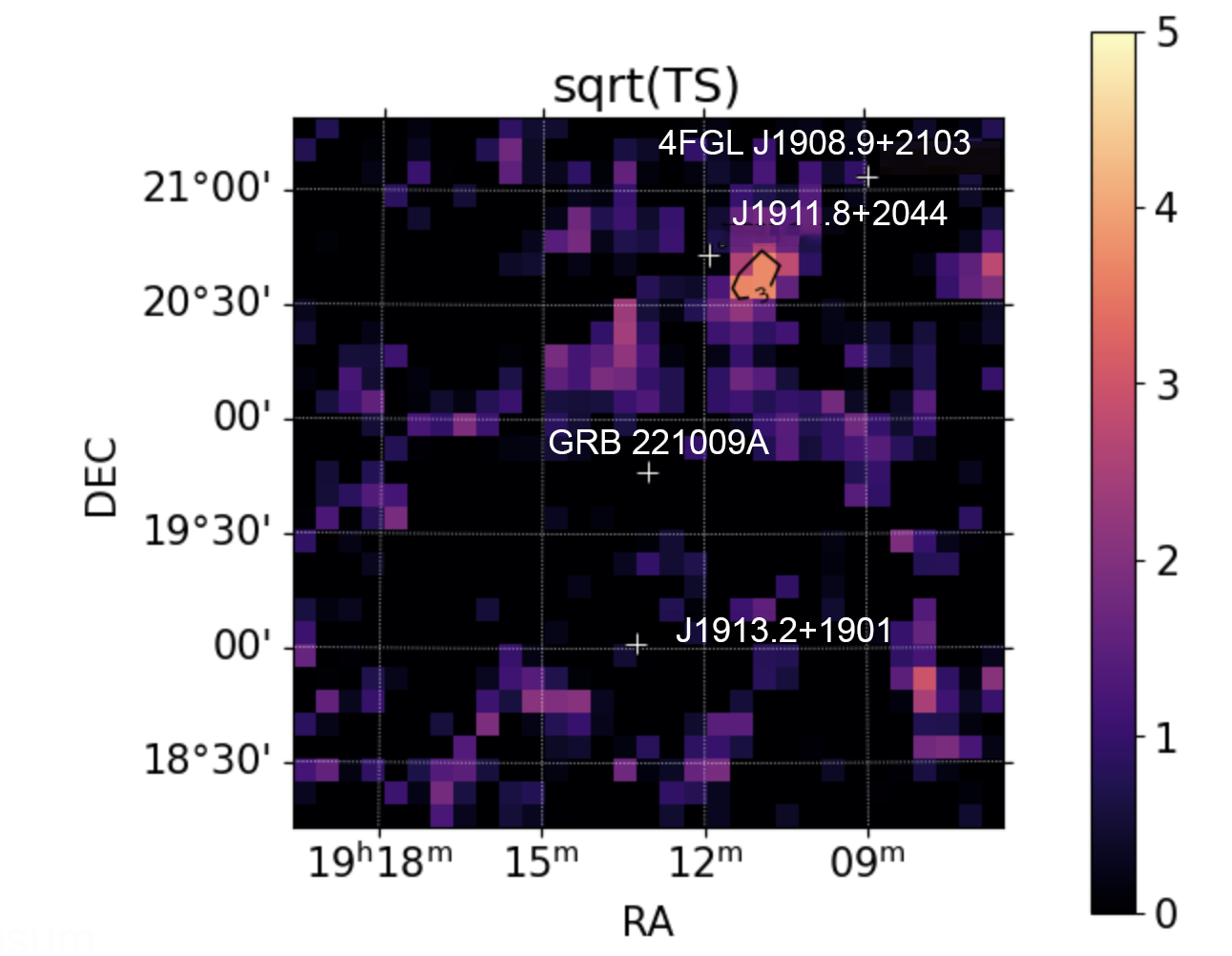}
    \includegraphics[width=0.48\textwidth]{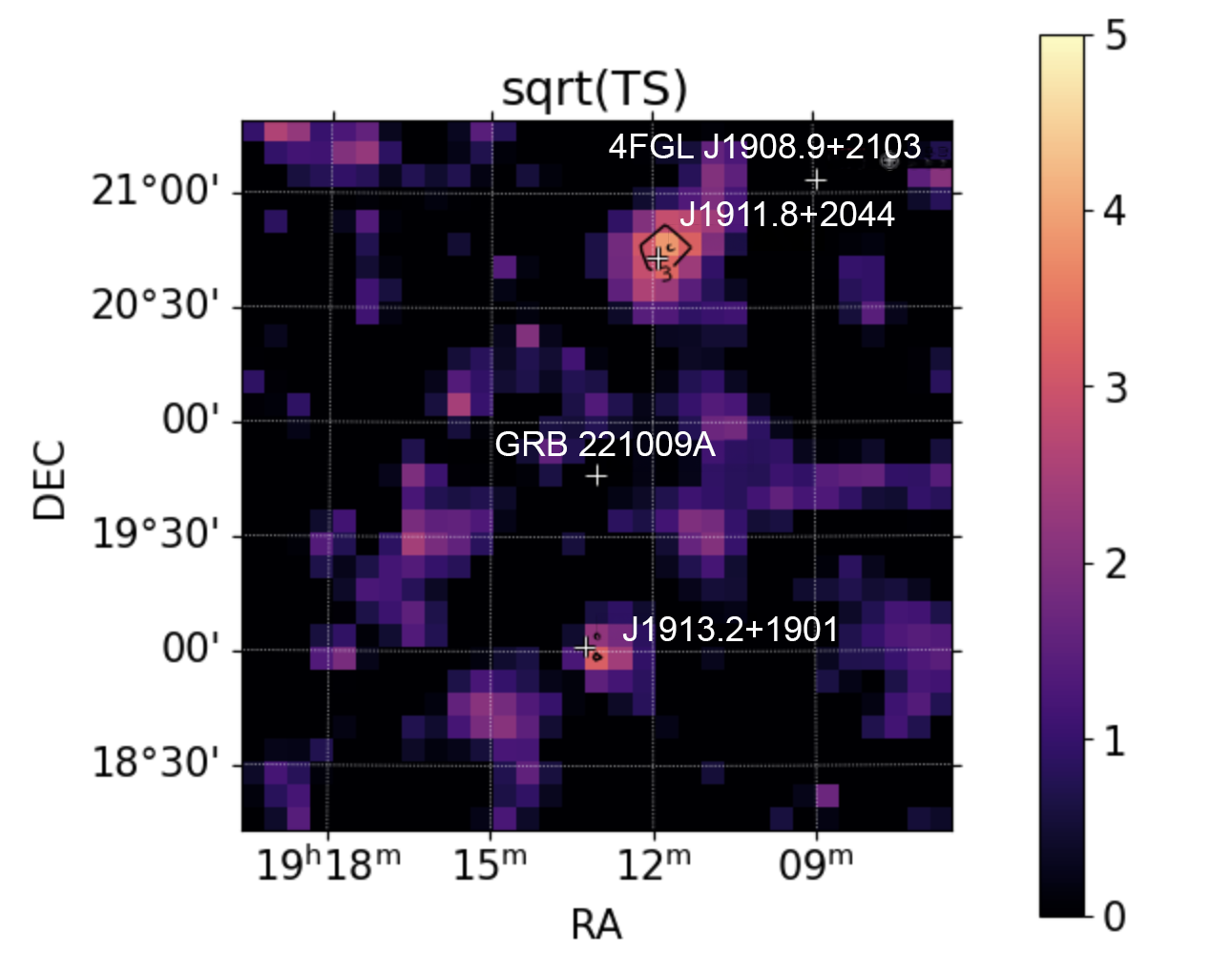}
    \caption{Left panel: TS  map (1 GeV -- 1 TeV) of the $3^{\circ} \times 3^{\circ}$  region centered on GRB~221009A using 14.2 years of pre-burst {\it Fermi}-LAT data. Right panel: TS  map generated using LAT data from 5.73 to 456.69 days after the GBM trigger for GRB~221009A. In both cases, J1911.8+2044 and J1913.2+1901 are excluded from the model. The white cross in the center of the figure marks the location of GRB~221009A.}
    \label{fig1}
\end{figure*}

We model a scenario in which UHE protons undergo $p\gamma$ interactions with the internal GRB photons. The UHE $\gamma$-rays at $\gtrsim 10^{19}$~eV produced during the prompt emission \citep{Razzaque:2004ApJ...613.1072R, Murase:2009ah} or the early afterglow phase \citep{Waxman:1999ai, Murase:2007yt, Razzaque:2013dsa} can escape from the strongly-magnetized GRB jet, avoiding significant $e^\pm$ pair production with synchrotron-self-absorbed photons, and are cascaded inside the structured region. The secondary $e^\pm$ pairs quickly lose their energy via synchrotron emission while they are strongly beamed. 
The resulting synchrotron pair echo emission \citep{Murase:2011yw}, 
uniquely predicts a time delay of $\mathcal{O}(100)$~days at $\varepsilon_\gamma\sim0.3-1$~TeV. From \cite{Murase:2011yw}, the characteristic time spread of the
synchrotron pair echo at energy $\varepsilon_{\gamma,\rm syn}$ is
\begin{equation}
\begin{split}
\Delta t_{\rm syn}&\approx \frac{{\rm min}[\lambda_{\gamma\gamma},d_c]\theta_{\rm syn}^2}{2c} \nonumber \\ &\simeq 100~{\rm days}~{\left(\frac{{\rm min}[\lambda_{\gamma\gamma},d_c]}{10~\rm Mpc}\right)} {\left(\frac{\varepsilon_{\gamma, \rm syn}}{1~\rm TeV}\right)}^{-2}.
\label{eqn:delay_syn}
\end{split}
\end{equation}
This timescale characterizes the
width of the delayed pair-echo signal, rather than the exact arrival
time of an individual photon. A delayed photon is therefore interpreted as lying within the
corresponding pair-echo transient window.

\begin{table*}
	\centering
	\caption{Sub-threshold Sources Detected with TS $> 9$}
	\label{table1}
	\begin{tabular}{lcccc} 
		\hline
		 Designation & RA & Dec. & TS & 95\% Flux UL (1 GeV--1 TeV)\\
         & degrees & degrees &  & erg cm$^{2}$ s$^{-1}$\\
		\hline
          J1911.8+2044 & 287.979 $\pm$ 0.070 & 20.720 $\pm$ 0.070 & 17.9 & $7.0 \times 10^{-12}$\\
		 J1913.2+1901 & 288.307 $\pm$ 0.120 & 19.011 $\pm$ 0.120 & 10.6 & $4.3 \times 10^{-12}$ \\
		\hline
	\end{tabular}
\end{table*}

The prompt GRB spectrum at high energies is modeled using a Band function above
the peak synchrotron energy $\varepsilon_{\gamma,\rm pk}$. The comoving photon
number density is estimated as
$n'_\gamma \simeq L_{\gamma,\rm iso}/(4\pi r_{\rm sh}^2 c \Gamma_b
\varepsilon_{\gamma,\rm pk})$ \citep{Razzaque:2004ApJ...613.1072R}, where
$L_{\gamma,\rm iso}\simeq10^{54}$ erg s$^{-1}$ is the isotropic-equivalent $\gamma$-ray luminosity,
$r_{\rm sh}=10^{14.5}\,{\rm cm}$ is the shock radius, $c$ is the speed of
light, and $\Gamma_b=10^{2.7}$ is the bulk Lorentz factor. These parameter
values are consistent with IceCube's non-detection of neutrinos from
GRB~221009A \citep{Murase:2022vqf}. Below and above the peak, the photon
spectrum is taken to scale as $\varepsilon_\gamma^{-1}$ and
$\varepsilon_\gamma^{-\alpha}$, respectively, with $\alpha>2$. The intrinsic UHECR spectrum is assumed to follow
$dN/dE^\prime_p\propto E_p^{\prime-2}$ over $10^{18}$--$10^{20.5}$ eV
\citep{Murase:2011yw}. The $\pi^0$-decay UHE $\gamma$-ray spectrum, resulting from $p\gamma$ interactions, ${dN_{\gamma, \pi}}/{d\varepsilon_{\gamma, \pi}}$ is calculated using the
parametrization of \citet{Kelner:2008ke}.

The produced UHE $\gamma$-rays are
assumed to escape the GRB emission region efficiently because the low-energy target photons
relevant for internal $\gamma\gamma$ absorption are suppressed by synchrotron
self-absorption \citep{Razzaque:2004ApJ...613.1072R}.
The synchrotron flux from $e^\pm$ pairs can be approximated as
\begin{equation}
\begin{split}
\varepsilon_{\gamma,\rm syn}^{2}{\Phi_{\gamma,\rm syn}}
\simeq
\frac{(1-e^{-{l_{\rm SR}}/{\lambda_{\gamma\gamma}}}
)}{4\pi d_L^2}
f_{\rm esc} \left(\varepsilon_{\gamma, \pi}^{2}\frac{dN_{\gamma, \pi}}{d\varepsilon_{\gamma, \pi}}\right) .
\end{split}
\end{equation}
where $l_{\rm SR}=10$ Mpc is the dimension of the structured region, and $\lambda_{\rm \gamma\gamma}$ is the $\gamma\gamma$ pair production optical depth in the CMB and EBL photon field at the redshift of the source.
The characteristic synchrotron photon energy at Earth is
\begin{equation}
\begin{split}
\varepsilon_{\gamma,\rm syn}
\simeq
0.12\,{\rm TeV}\, (1+z)^{-1}
\left(\frac{\gamma_{e}}{10^{13.5}}\right)^{2}
\left(\frac{B_{\rm SR}}{10\, \text{nG}}\right),
\end{split}
\end{equation}
The spectrum of synchrotron pair echo emission is shown in Fig.~\ref{fig:pair_echo}. The dotted line shows the spectrum at production, and the solid line is that after attenuation in the EBL. For the latter, we have used the EBL model of \citet{Gilmore_2012}. The vertical dashed line is the characteristic synchrotron photon energy corresponding to a time delay of $\approx192$ days. 
The pair deflection in the structured magnetic field is
\begin{equation}
\begin{split}
\theta_{\rm EG}
\simeq
\frac{\sqrt{2}\lambda_{\rm syn}}{\sqrt{3}r_L}
\simeq
0.063^\circ\,
\left(\frac{\gamma_{e}}{10^{13.5}}\right)^{-2}
\left(\frac{B_{\rm SR}}{10\, \text{nG}}\right)^{-1}.
\end{split}
\end{equation}
Thus, the pair-echo emission is highly beamed, although the observed angular offset can be broadened by the intrinsic jet
opening angle, making this a viable channel for delayed off-axis $\gamma$-ray
emission on timescales of $\lesssim1$ yr. We note that this scenario requires the escaping UHE $\gamma$ rays
to populate the jet cone on angular scales comparable to $\theta_j$. If instead, they
are confined to $\Gamma_b^{-1}\simeq0.1^\circ$, explaining the
observed $0.75^\circ$ offset becomes challenging.

\section{{\it Fermi}-LAT Search for Delayed Signal\label{sec:fermi}}


To study delayed $\gamma$-ray emission in the region around GRB~221009A, we used the \texttt{Fermitools} package (v.2.2.0) and \texttt{fermipy} package (v.1.4.0) to produce a test-statistic (TS) map, covering a region of interest (ROI) of $10^{\circ} \times 10^{\circ}$ with pixel size of $0.1^{\circ}$. The map is generated by adding a tentative point source at each location and then calculating the likelihood ratio between the previously optimized model and the best-fit model with the additional source at that location. The test statistic is defined as TS \(= -2~{\rm log} \left({\mathcal{L}_0}/{\mathcal{L}_s}\right)\), where $\mathcal{L}_0$ is the likelihood of the original model and $\mathcal{L}_s$ is the likelihood of the model with the additional source. To avoid contamination from the GRB, we used 1 GeV--1 TeV events from 5.73 to 456.69 days after the burst trigger \citep{Axelsson_2025}. We selected \texttt{P8R3\_SOURCE\_V3} events with event class \texttt{evclass = 128}/ FRONT+BACK event types (\texttt{evtype=3}) and maximum zenith angle cut at $90^{\circ}$ to minimize Earth limb contamination. Additional contributions from nearby point sources were modeled using the 4FGL-DR4 catalog and standard background templates for the Galactic diffuse emission {\tt (gll\_iem\_v07.fits)} and the isotropic component {\tt (iso\_P8R3\_SOURCE\_V3\_v1.txt)} within $15^{\circ}$ of the RoI center.

We generated TS maps by performing a likelihood ratio test over a grid of possible coordinates, assuming a point source with a simple power-law spectral index, $\Gamma = -2$. To find the best fit, we allowed the isotropic and Galactic diffuse background components as well as the sources with TS $\geq 10$ within $3^{\circ}$ from the RoI center to vary in normalization.  To find the most significant features, we used the iterative source-finding algorithm {\tt gta.find\_sources} in \texttt{fermipy} to locate emission peaks on the TS map.  The most prominent features ($9 \leq TS \leq 25$) correspond to two sub-threshold sources J1911.8+2044 and J1913.2+1901\footnote{We consider sub-threshold sources to be those observed with less than $5\sigma$ significance and not present in existing catalogs} (see Fig.~\ref{fig1}). As a final step, we used the {\tt gta.localize} \texttt{fermipy} subroutine to refine the position of each residual (see Table \ref{table1}). The listed uncertainty corresponds to a 68\% C.L. radius of the error ellipse. Fig.~\ref{fig1} shows $3^{\circ} \times 3^{\circ}$ pre-burst and post-burst TS maps centered on the GRB~221009A location after excluding J1911.8+2044 and J1913.2+1901 from the model. One can see that in both cases, the field is textured with several structures in the TS map.

The most prominent sub-threshold peak in the TS map J1911.8+2044 is offset by $0.983^{\circ}$ from the GRB~221009A position. $\gamma$-ray emission in this region appears to peak after the GRB trigger. However, as can be seen in the pre-burst TS map (Left Panel in Fig.~\ref{fig1}), there appears to be $\gamma$-ray emission in the proximity of J1911.8+2044 at RA = 287$^{\circ}.764$ $\pm$   0.053, Dec = 20.652 $\pm$ 0.054. Since we cannot unequivocally rule out $\gamma$-rays potentially produced before the GRB onset at this location, we discard the possibility that GRB~221009A induced J1911.8+2044.

The other sub-threshold residual of interest is J1913.2+1901, which is located $0^{\circ}.75$ from GRB~221009A. It also coincides with a 664.6 GeV event that arrived approximately 191.9 days after the GBM trigger. The total 95\% PSF containment radius for a 664.6 GeV event is about $0.3^{\circ}$. Thus, the residual appears to be disjointed from GRB~221009A and is most likely outside the host galaxy.



In our model, the brightest sub-threshold peak is used to derive an upper limit on secondary gamma-ray emission from GRB 221009A in the energy range $0.1-1$ TeV. Assuming a power law with photon index $\Gamma=-2$, the LAT upper limit (95\% confidence level, 100 GeV -- 1 TeV) is $\leq 2.3 \times 10^{-11}$ erg~cm$^{-2}$~s$^{-1}$ over 5.73-456.69 days after the GBM trigger.  We choose to provide 95\% confidence level upper limits throughout the paper since the TS $\leq$ 25. Analysis of the same region 14.2 years prior to GRB~221009A indicates a pre-burst flux upper limit $\leq 1.5  \times 10^{-12}$ erg cm$^{-2}$ s$^{-1}$ at the same location and in the energy range $0.1 - 1$ TeV. We proceed by investigating theoretical models for secondary emission induced by GRB 221009A that are consistent with the upper limit derived from post-event {\it Fermi} LAT \(\gamma \)-ray observations.

\section{Modeling~GRB 221009A early afterglow and delayed spectra}
\label{subsec:broadband}

To apply the models in Sec.~\ref{sec:models}, we use the LAT upper limit to constrain such a signal from GRB~221009A.
VHE emission detected from GRBs \citep{Abdalla2019_GRB180720b, Abdalla2021_GRB190829a, MAGIC2019Natur575455M} are typically modeled using SSC emission by relativistic electrons in external forward shock ~\citep[e.g.,][]{2019ApJ...880L..27D, 2021ApJ...923..135D, Wang2019ApJ...884..117W, 2021MNRAS.504.5647S, Joshi:2019opd, yama2022MNRAS142Y, Salafia2022, Sato:2022wec, Barnard:2024MNRAS.52711893B, Barnard2025MNRAS.543.4218B}, including the TeV emission detected by LHAASO from GRB~221009A \citep{LHAASO:2023kyg, Zhang:2022lff, Zhang:2023uei, Das:2022gon, 2024ApJ...962..115R, Banerjee:2024hxp, 2023MNRAS.522L..56S, Sato:2025xjz, Barnard2025MNRAS.543.4218B}. 

\subsection{Broadband emission from GRB~221009A}

\begin{figure}
    \centering
    \includegraphics[width=0.46\textwidth]{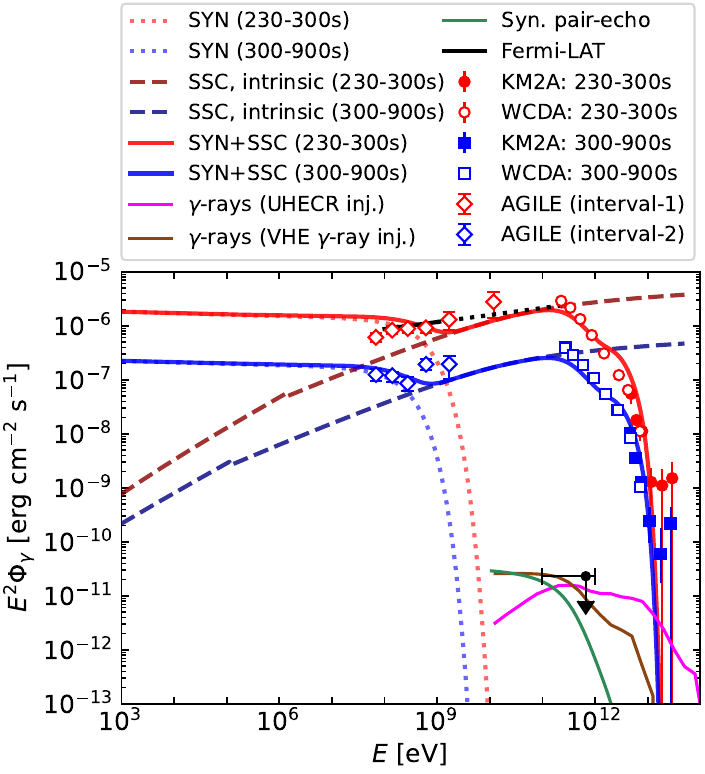}
    \caption{LHAASO data \citep{LHAASO:2023kyg} and {\it Fermi}-LAT detection \citep{Pillera_2022} in the 0.1-1 GeV range are shown. The filled black circle with a downward arrow represents the upper limit derived by {\it Fermi}-LAT for the 664.6 GeV photon at $0^\circ.75$ offset 191.9 days post-trigger. The observed flux by the AGILE detector has been shown for 248--326 s (interval-1) and 326-980 s (interval-2) \citep{Tavani2023ApJ...956L..23T}. 
    The synchrotron and SSC models are shown as dotted and dashed curves, respectively, while the solid curves correspond to the total, including EBL attenuation.
    The magenta and brown curves show UHE proton-induced cascade $\gamma$-rays from the structured region and intergalactic regions, respectively. The green curve shows the synchrotron pair-echo emission.
    }
    
    
    \label{fig:uhecr_gamma}
\end{figure}

In Fig.~\ref{fig:uhecr_gamma}, we show the spectrum of GRB~221009A for our synchrotron and SSC emission model during 230-300~s and 300-900~s after the {\it Fermi}-GBM trigger, assuming that the blastwave is already in decelerating phase.
The red- and blue-dotted curves indicate the synchrotron spectrum during the time intervals $T_0+(230 - 300)$~s and $T_0+(300 - 900)$~s, respectively. The SSC fluxes, fitted to LHAASO data for the two intervals \citep{LHAASO:2023kyg} and to AGILE data during 248--326 s (interval-1) and 326-980 s (interval-2) \citep{Tavani2023ApJ...956L..23T}, are shown by red-dashed and blue-dashed curves. The red and blue solid curves represent the total leptonic emission with EBL absorption using the \cite{Finke10} model. 
The synchrotron and SSC model curves were calculated following \cite{Joshi:2019opd, Barnard:2024MNRAS.52711893B, Das:2022gon, Huang2026EPJC...86..195H} with best-fit parameters obtained by \citep{Huang2026EPJC...86..195H} as $E_k = 10^{55}$ erg, $n_0 = 1$~cm$^{-3}$, $p = 2.05$, $\epsilon_e = 2\times 10^{-3}$ and $\epsilon_B = 6\times 10^{-4}$. It should be noted that a different set of parameter values can also yield a satisfactory fit to the data due to parameter degeneracy.
Moreover, the SSC spectrum lies in the Thomson regime, and has a break at $\approx 0.5$~TeV due to the Klein-Nishina effect~\citep{Nakar:2009er, Joshi:2019opd}. The spectrum is therefore, $dn/d\varepsilon_\gamma \propto \varepsilon_\gamma^{-1.5}$ for $\varepsilon_\gamma \lesssim 1$~TeV and $\propto \varepsilon_\gamma^{-p} \propto \varepsilon_\gamma^{-2.05}$ for $\varepsilon_\gamma \gtrsim 1$ ~TeV. As an approximation, we consider $p=2$ for the $\gamma$-ray injection in our VHE $\gamma$-ray injection model.

While VHE emission detected from GRBs is dominated by SSC in the forward shock of a GRB blast wave, the SSC flux typically declines as $F_\nu \propto t^{\alpha}$. In the case of a constant-density interstellar medium (ISM), we have $\alpha = -(9p -10)/8$, and in the case of a wind-type medium $\alpha = -p+1$, both for the observed emission above the peak of the SSC spectrum in the Thomson regime \citep[see, e.g.,][]{Joshi:2019opd}. For $p \approx 2$, the index for the electron energy distribution is $\alpha \approx -1$. The flux decay observed by LHAASO in the $0.2-5$~TeV range is $\alpha_{2,\rm TeV} = -1.115\pm 0.012$ in their second time interval of $\approx T_0 + 230$~s to $\approx T_0 + 670$~s~\citep{LHAASO:2023kyg}. However, LHAASO also observed a further steepening of the flux at $\gtrsim T_0 + 670$~s with a flux-decay index $\alpha_{3,\rm TeV} = -2.21^{+0.30}_{-0.83}$. If interpreted as the post-jet-break decay \citep{Sari1999ApJ...519L..17S}, for which $F_\nu \propto t^{-p}$ and for $p \approx \alpha_{3,\rm TeV}$, the $0.2-5$~TeV flux at $T_0 + 191.9$ day, as an example, would be $\approx 3\times 10^{-7}(191.9~{\rm day}/670~{\rm sec})^{\alpha_{3,\rm TeV}} \sim (3\times 10^{-13} - 3\times 10^{-18})$ erg cm$^{-2}$~s$^{-1}$, which is well below the {\it Fermi}-LAT flux upper limit shown in Fig.~\ref{fig:uhecr_gamma} and cannot explain the 664.6~GeV photon.
Thus, we explore various delayed off-axis emission scenarios, described in Sec.~\ref{sec:models}, as possible origins of this photon.

\subsection{Energy Requirement in primary UHECR and $\gamma$-ray injection\label{subsec:budget}}


Fig.~\ref{fig:uhecr_gamma} shows the $\gamma$-ray flux at Earth from primary UHE proton, primary VHE $\gamma$-ray injection, and synchrotron pair echo emission due to primary UHE $\gamma$-ray injection by the magenta, brown curves, and green curves, respectively, normalized to the flux upper limit of the 664.6 GeV photon at an angular offset of $0^\circ.75$. We adopt the upper limit on the sub-threshold feature as a fiducial flux to test our theoretical models. In the 3-D simulations described in Sec.~\ref{sec:models}, events are selected within a $1^\circ$ aperture at the edge of the galaxy cluster, and the same angular cut is applied after propagation through the IGM for events collected at Earth. Given the $0^\circ.75$ angular offset of the sub-threshold photon and its 95\% PSF containment radius of $0^\circ.3$, adopting a $1^\circ$ aperture provides a reasonable approximation. 

The fractions of the injected primary UHECR and VHE $\gamma$-ray energy transferred to the final cosmogenic $\gamma$-ray spectrum within the viewing aperture above $\gtrsim 10$ GeV are $f_{\gamma,p}=3.3\times10^{-3}$ and $f_{\gamma,\gamma}=0.33$, respectively. In addition, among the events within the viewing aperture at the edge of the galaxy cluster, the fraction with momentum vectors within $1^\circ$ of the initial emission axis is $\xi^p_{B,{\rm SR}}(\leq1^\circ)=0.015$ for secondary protons (see Fig.~\ref{fig:uhecr_defl}) and $\xi^\gamma_{B,{\rm SR}}(\leq1^\circ)\simeq 1$ for photon injection (see Fig.~\ref{fig:gamma_defl}). For computational efficiency, we do not further isolate the subset of events
whose momentum vectors exactly match those of the sub-threshold photons observed
at Earth. 
Instead, we use the total flux within the aperture cone, which provides a
conservative lower limit on the source energy requirement.

The following expression yields the required energy at injection, considering a steady cosmogenic $\gamma$-ray spectrum,
\begin{equation}
\begin{split}
E \gtrsim \dfrac{2\pi d_L^2 (1-\cos\theta_j)}{\eta(\leq1^\circ)} \Delta T\int\varepsilon_\gamma \dfrac{dn}{d\varepsilon_\gamma dA dt} d\varepsilon_\gamma 
\label{eqn:lum}
\end{split}
\end{equation}
where $\eta(\leq1^\circ)=\xi^p_{B, \rm SR}(\leq1^\circ) \times f_{\gamma, p}$, for UHE proton injection model and $\eta(\leq1^\circ)=\xi^\gamma_{B, \rm SR}(\leq1^\circ) \times f_{\gamma, \gamma}$ for VHE $\gamma$-ray injection model, $\Delta T=191.9$ days is the time delay of the 664.6 GeV photon, and $\varepsilon_\gamma$ ranges over the entire $\gamma$-ray spectrum above 10 GeV. The proton injection time can be much shorter than $T$. Here, the integration is over the cosmogenic $\gamma$-ray flux. Eqn.~\ref{eqn:lum} gives $E_p \gtrsim 8.3\times 10^{52}$ erg and $E_\gamma\gtrsim1.7\times10^{49}$ erg, for proton and $\gamma$-ray injections respectively. The corresponding isotropic equivalent energy is given as $E^{\rm iso}_p\gtrsim 1.7\times 10^{57}$ erg and $E^{\rm iso}_\gamma\gtrsim3.4\times10^{53}$ erg.

 The isotropic-equivalent VHE $\gamma$-ray luminosity from broadband modeling of the GRB jet is (cf. Sec.~\ref{subsec:broadband}), 
\begin{equation}
\begin{split}
L_{\rm VHE\gamma}^{\rm iso}= 4\pi d_L^2 \int d\varepsilon_\gamma \varepsilon_\gamma \frac{dn}{d\varepsilon_\gamma} \sim 2\times 10^{50} \text{\ erg s}^{-1}. 
\label{eqn:lvhe}
\end{split}
\end{equation}
Now, assuming a duration of $\sim1000$~s, the isotropic-equivalent energy is $E_{\rm VHE\gamma}^{\rm iso}\sim2\times 10^{53}$~erg. Assuming a constant $t^{-1}$ decay as expected in the SSC model, the VHE $\gamma$-ray luminosity at a time given in Eq.~\ref{eqn:delay_syn}
is estimated to be $\sim 10^{46}~\text{\rm erg s}^{-1}$. 
Note that in the case of a $t^{-2}$ decay, as expected from the jet-break detected by LHAASO, the flux $\mathcal{O}(100)$~days after the burst would be too low. 
Thus, the GRB afterglow scenario alone is not viable for the origin of any delayed emission in the latter case, and late-time injection is necessary if VHE emission is attributed to the GRB jet. For our UHE $\gamma$-ray injection scenario, an assumed value of $L_{\gamma, \rm iso}\sim 10^{54}$ erg s$^{-1}$ during the prompt emission phase yields $E_{p, \rm iso}\simeq 2.8\times10^{47}(\Delta t_{\rm syn}/T_{\rm dur})$ erg 
in the $10^{18}$--$10^{20.5}$ eV range. Assuming a pair-echo time spread of $\Delta t_{\rm syn}\simeq200$ days and prompt emission duration of $T_{\rm dur}=20$~s, the isotropic equivalent proton energy required is $L_{p,\rm iso}=2.3\times10^{53}$ erg.

\subsection{Testing the Sub-threshold Feature Against the Models}

The UHECR and VHE $\gamma$-ray injection models can account for the
$0^\circ.75$ angular offset of the sub-threshold photon flux upper limit.
The UHE $\gamma$-ray injection scenario is more constrained, since the
secondary pair deflection is small and the escaping UHE photons may be
confined to the relativistic beaming angle, rather than the full jet cone.
But the transverse distance corresponding to $0^\circ.75$ angular offset at redshift $z=0.151$ corresponding to GRB~221009A is $\approx8.5$ Mpc. Hence, the geometric time delay from Eqn.~\ref{eqn:time_geo} is $2.8 \times 10^7$ yrs, which is much larger than that of protons in the structured region (see Eq. \ref{eqn:delay_igm} and also Fig.~\ref{fig:gamma_delay}. Hence, it cannot be reconciled with the timing of the sub-threshold $\gamma$-ray photon. Nevertheless, we use the subthreshold photon as an upper limit in our model to compute the cosmogenic $\gamma$-ray flux distribution from magnetized structure and the required source energetics, which has broader implications for other off-axis sources and for orphan neutrino, $\gamma$-ray, and cosmic-ray observations without multimessenger counterparts \citep{TelescopeArray:2023sbd, KM3NeT:2025ccp, Das:2025tfq, Das:2025vqd, Sherman:2025gir, Boxi:2025ony}. 

For UHE proton injection, $\gamma_e\approx10^{9}$ for $e^\pm$ produced by UHECRs at $\approx1$ EeV, the lowest energy in our simulations, due to the BH process and an inelasticity of $\approx10^{-3}$. This gives a time delay of the order of seconds for a pencil beam, but UHECRs undergo a time delay of $\sim10^3$ years in the galaxy cluster region (see Fig.~\ref{fig:uhecr_delay}).
In addition, as shown in the previous section, the required energy in the UHE proton-induced $\gamma$-ray scenario is too demanding for the time delay of our example sub-threshold upper limit.

For VHE $\gamma$-ray injection, $\gamma_e \sim 3\times 10^{7}$ is required to upscatter a CMB photon to 665 GeV via IC emission. This corresponds to a time delay of $\sim 1.5$~yrs for a pencil beam, with a magnetic field $B_{\rm IGM}\approx10^{-5}$ nG (see Sec.~\ref{subsec:uhecr}).
The VHE $\gamma$-ray induced cascade emission poses a more energetically acceptable alternative. Eq.~\ref{eqn:lvhe} implies the injected isotropic-equivalent VHE $\gamma$-ray energy is $E_{\rm UHE\gamma}^{\rm iso}\sim{10}^{53}$~erg for an emission duration of $\sim1000$~s, which is only $\sim1$\% of the observed total isotropic-equivalent $\gamma$-ray energy. This corresponds to a synchrotron luminosity of $\sim10^{46},{\rm erg,s^{-1}}$ at $\sim0.3$–$1$~TeV, assuming a post-event $t^{-1}$ flux decay. This level is compatible with the detection of the 664.6~GeV photon. Although the VHE $\gamma$-ray–induced cascade is energetically more favorable, the geometric time delay 
still disfavor an origin in this GRB.

For the UHE $\gamma$-ray injection scenario, we assume that UHECRs are accelerated in internal shocks during the prompt emission. The
non-detection of astrophysical neutrinos from GRB~221009A by IceCube severely constrains the allowed model parameters. Within this range, synchrotron pair-echo emission can produce a significant delayed flux on timescales
$\lesssim1$ yr. 
If normalized close to the median energy of the delayed photon, the model predicts
a larger GeV-band flux. However, within the energy uncertainty, the flux is consistent with the derived upper limit. The spectrum becomes flat at lower energy (see Fig.~\ref{fig:pair_echo}).
The emission is strongly beamed, with typical deflections
$\lesssim\theta_j$. The jet opening angle can provide additional angular
broadening, allowing the emission to fall within the angular uncertainty of the
late photon. The required UHECR energy is $\lesssim1\%$ of the
isotropic-equivalent $\gamma$-ray energy emitted during the prompt phase.

To conclude, although our models may reproduce the angular offset of the sub-threshold photon, the geometric time delay cannot be reconciled with an origin in this GRB.

\subsection{Calculation of the expected number of events\label{sec:event_rate}}

\begin{table}
\centering
\caption{Poisson probability of observing N$_{\rm obs}$ events given N$_{\rm exp}$ expected events in four different energy bands and for four different time-interval models.}
    \label{tab:my_label}
    
    \begin{tabular}{rrrr}
         Energy Band & N$_{\rm obs}$ & N$_{\rm exp}$ & P(N$_{\rm obs}$|N$_{\rm exp}$) \\
        \hline
        \multicolumn{4}{c}{Model a: T$_{0}$+5.73~d to T$_{0}$+457.0~d, $\Gamma$=-2.5}\\
        \hline
         1 GeV   -- 1 TeV & 716 & 865.02  & $<10^{-5}$  \\
         10 GeV  -- 1 TeV & 24  & 32.6  & $2.8\times10^{-4}$  \\
         100 GeV -- 1 TeV &  1  & 1     & $3.7\times10^{-1}$ \\
         600 GeV -- 1 TeV &  1  & 0.035 & $3.4\times10^{-2}$ \\
        \hline
        \multicolumn{4}{c}{Model b: T$_{0}$+5.73~d to T$_{0}$+457.0~d, $\Gamma$=-2}\\
        \hline
         1 GeV   -- 1 TeV & 716 & 86.5  & $<10^{-5}$  \\
         10 GeV  -- 1 TeV & 24  & 11.1  & $2.8\times10^{-4}$  \\
         100 GeV -- 1 TeV &  1  & 1     & $3.7\times10^{-1}$ \\
         600 GeV -- 1 TeV &  1  & 0.069 & $6.6\times10^{-2}$ \\
         \hline
        \multicolumn{4}{c}{Model c: T$_{0}$+5.73~d to T$_{0}$+457.0~d, $\Gamma_1$=-1.8, $\Gamma_2$=-2.3}\\
        \hline
         1 GeV   -- 1 TeV & 716 & 56.1  & $<10^{-5}$  \\
         10 GeV  -- 1 TeV & 24  & 10.5  & $7.7\times10^{-5}$  \\
         100 GeV -- 1 TeV &  1  & 1     & $3.7\times10^{-1}$ \\
         600 GeV -- 1 TeV &  1  & 0.047 & $4.5\times10^{-2}$ \\
        \hline
             \multicolumn{4}{c}{Model d: T$_{0}$+100~d to T$_{0}$+457.0~d, $\Gamma_1$=-1.8, $\Gamma_2$=-2.3}\\
         \hline
          1 GeV   -- 1 TeV & 576 & 57.1  & $<10^{-5}$  \\
          10 GeV  -- 1 TeV & 22  & 10.1  & $4.4\times10^{-4}$  \\
          100 GeV -- 1 TeV &  1  & 1     & $3.7\times10^{-1}$ \\
          600 GeV -- 1 TeV &  1  & 0.047 & $4.5\times10^{-2}$ \\
          \hline        
\end{tabular}
    
\end{table}

To ensure that our {theoretical} models do not over-predict the observed data, we performed a cross-check by assuming a simple power-law model $dn/d\varepsilon_\gamma\propto \varepsilon_\gamma^{\Gamma}$ and calculating the exposure in four energy bands (from 1--, 10--, 100--, and 600~GeV-- to 1~TeV) during the time period \( T_0+5.73 \) days to \( T_0+457.0 \) days, for spectral indices $\Gamma = -2.5$ (Model a) and $\Gamma = -2.0$ (Model b). For each case, we determined the flux required to produce exactly one expected event between 100 GeV and 1 TeV, and then we computed the expected number of events (\( N_\text{exp} \)) in all four energy bands. These results were compared with the observed number of events (\( N_\text{obs} \)). The table below summarizes \( N_\text{obs} \), \( N_\text{exp} \), and the probability of observing \( N_\text{obs} \) given \( N_\text{exp} \).

We also tested a broken power-law model (\( \Gamma_1 = -1.8 \) for \( E < 140 \, \mathrm{GeV} \), \( \Gamma_2 = -2.3 \) for \( E \geq 140 \, \mathrm{GeV} \); Model c), representative of the cascade model in Fig.~\ref{fig:uhecr_gamma}, and the same model, but calculating the exposure between \( T_0+100 \) days to \( T_0+457.0 \) days (Model d). A single power law with \( \Gamma = -2.0 \) was consistent with the observations, while softer indices (\( \Gamma > -2.0 \)) overpredicted the counts below 100 GeV. The broken power-law model also agreed well with the data, predicting fewer events below 100 GeV while maintaining a reasonable probability of observing events above 600 GeV. Very similar results are obtained when computing the exposure starting from 100 days after the trigger, 
which is a proxy for cascade flux increasing sharply with time.

\section{Discussions and Summary\label{sec:discussion}}
We develop a framework to test delayed off-axis emission from $\gamma$-ray emitters embedded in large-scale magnetized structures. For a strong magnetic field, $\mathcal{O}(10~\mathrm{nG})$, in the structured
region, a fully analytical treatment is complicated by the finite opening angle
of the injected beam, broad injection spectrum, and coupled cascade processes.
We perform 3-D numerical simulations to derive time delays and deflection of $\gamma$-ray signals in various
scenarios. We apply the models to GRB~221009A as a representative case embedded
in a galaxy cluster. The energy requirement for delayed off-axis emission is
estimated using flux upper limits derived from the Fermi-LAT observation of a sub-threshold photon of energy 664.6 GeV, at $0^\circ.75$ offset from GRB 221009A, $\sim 191.9$ days post-trigger. 
Although the angular offset can be explained, the geometric time delay disfavors an origin of this sub-threshold feature in this GRB.

Our two-zone magnetic-field framework provides a basis for studying orphan neutrino, $\gamma$-ray, and cosmic-ray signals lacking prompt multimessenger counterparts. \citep{TelescopeArray:2023sbd, KM3NeT:2025ccp, Das:2025tfq, Das:2025vqd, Sherman:2025gir, Boxi:2025ony}. Such signals can arise from delayed off-axis emission when the source is embedded in a magnetized structure. The delay distributions in Figs.~\ref{fig:uhecr_delay} and \ref{fig:gamma_delay} should not be interpreted as the duration over which the cascade emission would appear as a source-resolved LAT counterpart. Although GRB~221009A-like events have been estimated to occur at rates ranging from once per $\sim10^4$ yr \citep{2023ApJ...946L..31B} to once per $\sim200$ yr for a possible narrow-jet population \citep{Finke:2024mni}, the delayed secondary cascade photons produced in the magnetized structured region are strongly isotropized and are therefore more naturally expected to contribute to the Fermi-LAT isotropic diffuse $\gamma$-ray background \citep{Fermi-LAT:2014ryh}. Thus, identifying GRB cascade emission among unassociated LAT sources \citep{Ballet:2026kpq} would require a population-level study of the surrounding sky regions of past transients.

Our main results are summarized
as follows.

\begin{enumerate}[leftmargin=*]
\setlength{\itemsep}{0em}
    \item UHECRs, injected by the GRB jet, can propagate through structured regions of the IGM and generate secondary cascade emission via interactions with EBL and CMB photons. However, the IC cascade in the structured region is severely diluted. Deflected UHECRs escaping the magnetized structure can produce off-axis $\gamma$-ray upon further propagation in the void IGM. The geometric delay is of the order tens of Myr, while the UHE protons experience a time delay of $\sim$kyr within the cluster region. 
    This model requires a weaker magnetic field in the IGM, $B_{\rm IGM}\sim 10^{-7}$ nG. The isotropic-equivalent energy in UHE protons required to power the cascade emission is $\sim 10^{57}\,{\rm erg}$. This corresponds to a substantially high energy released in UHE protons over the delay time scale of the sub-threshold photon, approximately $\sim 100$ times that compared to the isotropic-equivalent energy released in $\gamma$-rays  \citep{Axelsson_2025}.

    \item In the second scenario, EBL-unattenuated VHE $\gamma$-rays produced during afterglow are injected. 
    The required power at source for the cascade signal at Earth is less demanding than the UHECR injection case, which corresponds to only $\sim1$\% of the isotropic-equivalent $\gamma$-ray energy and may explain the angular offset, but the geometric time delay disfavors an origin of the sub-threshold photon in this GRB. The required magnetic field in the void IGM in this case is $B_{\rm IGM}\sim 10^{-5}$ nG to account for additional deflection of $e^\pm$ secondaries in the IGM.

    \item In the UHE $\gamma$-ray injection scenario, UHECRs accelerated during the GRB prompt emission phase interact with the internal GRB photon field, producing escaping UHE $\gamma$ rays. These photons cascade in the surrounding magnetized structure, where secondary pairs emit synchrotron radiation efficiently. The resulting pair echo is strongly beamed and can produce delayed TeV emission on timescales of $\lesssim1$ yr. 
    Within the parameter range consistent with IceCube's non-detection of neutrinos from GRB~221009A, 
    the energy requirement in this model is $\lesssim1\%$ of the isotropic-equivalent prompt $\gamma$-ray energy.
    
    
\item An independent check on the expected number of events between 0.1-1 TeV, 5.73 days after the burst was performed, using various spectral shapes (power-law and broken power-law) and time intervals (5.73 days and 100 days after the burst). We found that the UHE proton-induced or UHE/VHE $\gamma$-ray-induced cascade flux, approximated by the broken power laws, does not over-produce VHE events in the LAT.


\item For a fixed magnetic field in the structured region, a larger jet opening angle of a few degrees, as inferred from radio and X-ray polarization \citep{Bright:2023izk, RenJia:2023uzb, OConnor:2023ieu, Negro:2023cer}, results in larger angular dispersion of UHECRs from the GRB. For our energy estimate, we compute the proton fraction within $0.9^\circ$ of the initial emission direction, to account for the Fermi-LAT resolution of $0.15^\circ$ above 10 GeV. A larger $\theta_j$ reduces this fraction, thereby increasing the required proton energy.

\item Our results have implications for the observation of VHE $\gamma$-rays from GRBs other than GRB~221009A. We provide detailed angular and time-delay distributions of the EM cascade emission induced by primary UHECRs and VHE $\gamma$-rays. Future searches for delayed photons can be conducted using {\it Fermi}-LAT data and ground-based $\gamma$-ray detectors after GRB triggers. A stacking analysis across different GRB directions could test for a population-level excess of delayed emission. Our modeling can be used to search for UHECR signatures via these delayed photons and to explore the GRB environment. Alternatively, the detection of a steady flux of multi-TeV $\gamma$-rays without an associated neutrino or other multimessenger counterparts may indicate delayed emission from a past cosmological transient.

\end{enumerate}


\section*{Acknowledgments} 
The \textit{Fermi}-LAT Collaboration acknowledges generous ongoing support from a number of agencies and institutes that have supported both the development and the operation of the LAT as well as scientific data analysis. These include the National Aeronautics and Space Administration and the Department of Energy in the United States, the Commissariat \`a l'Energie Atomique and the Centre National de la Recherche Scientifique / Institut National de Physique Nucl\'eaire et de Physique des Particules in France, the Agenzia Spaziale Italiana and the Istituto Nazionale di Fisica Nucleare in Italy, the Ministry of Education, Culture, Sports, Science and Technology (MEXT), High Energy Accelerator Research Organization (KEK) and Japan Aerospace Exploration Agency (JAXA) in Japan, and the K.~A.~Wallenberg Foundation, the Swedish Research Council and the Swedish National Space Board in Sweden.

We thank Guillem Martí-Devesa for carefully reading the manuscript and for providing helpful comments. We also thank Jean Ballet for valuable discussions.  This research by S.D. and K.M. is supported by KAKENHI No.~20H05852. The work of K.M. is supported by the NSF Grants No.~AST-2108466, No.~AST-2108467, and No.~2308021. The material is based upon work supported by NASA under award number 80GSFC24M0006. S.R.\ was partially supported by a BRICS STI grant and by a NITheCS grant from the National Research Foundation, South Africa. This research has made use of data obtained through the High Energy Astrophysics Science Archive Research Center Online Service, provided by the NASA/Goddard Space Flight Center. Numerical computation in this work was partly carried out at the Yukawa Institute Computer Facility, Kyoto University, Japan.

\software{Fermipy \citep{2017ICRC...35..824W}, Fermitools \citep{2019ascl.soft05011F}, CRPropa3 \citep{AlvesBatista:2016vpy, AlvesBatista:2022vem}}

\bibliography{main_v3}{}
\bibliographystyle{aasjournal}



\end{document}